\documentclass[12pt,preprint]{aastex6}

\usepackage[]{natbib}
\usepackage{graphicx}
\usepackage{amssymb}
\usepackage{amsmath}
\usepackage{color}
\usepackage{ulem}

\newcommand{\gcas}{$\gamma$~Cas}

\input{macro.inp}

\slugcomment{Accepted for Publication in Astrophysical Journal}

\shorttitle{Variable Absorbers within $\gamma$ Cassiopeiae}
\shortauthors{Hamaguchi et al.}

\begin{document}

\title{Discovery of Rapidly Moving Partial X-ray Absorbers within $\gamma$ Cassiopeiae}

\author{K. Hamaguchi\altaffilmark{1, 2}, L. Oskinova\altaffilmark{3}, C. M. P. Russell\altaffilmark{4}, R. Petre\altaffilmark{4}, T. Enoto\altaffilmark{5,6}, K. Morihana\altaffilmark{7}, M. Ishida\altaffilmark{8}}

\altaffiltext{1}{CRESST and X-ray Astrophysics Laboratory NASA/GSFC, Greenbelt, MD 20771, USA, Kenji.Hamaguchi@nasa.gov}
\altaffiltext{2}{Department of Physics, University of Maryland, Baltimore County, 1000 Hilltop Circle, Baltimore, MD 21250, USA}
\altaffiltext{3}{Institute of Physics and Astronomy, University of Potsdam, 14476 Potsdam, Germany}
\altaffiltext{4}{X-ray Astrophysics Laboratory NASA/GSFC, Greenbelt, MD 20771, USA}
\altaffiltext{5}{The Hakubi Center for Advanced Research, Kyoto University, Kyoto 606-8302, Japan}
\altaffiltext{6}{Department of Astronomy, Kyoto University, Kitashirakawa- Oiwake-cho, Sakyo-ku, Kyoto 606-8502, Japan}
\altaffiltext{7}{Nishi-Harima Astronomical Observatory, Center for Astronomy, University of Hyogo, 407-2, Nichigaichi, Sayo-cho, Sayo, Hyogo, 670-5313, Japan}
\altaffiltext{8}{The Institute of Space and Astronautical Science, Japan Aerospace Exploration Agency, 3-1-1 Yoshinodai, Chuo-ku, Sagamihara, 252-5210, Japan}

\begin{abstract}
Gamma Cassiopeiae is an enigmatic Be star with unusually strong hard X-ray emission.
The \SUZAKU\ observatory detected six rapid X-ray spectral hardening events called ``softness dips" in a $\sim$100~ksec duration observation in 2011.
All the softness dip events show symmetric softness ratio variations,
and some of them have flat bottoms apparently due to saturation.
The softness dip spectra are best described by 
either $\sim$40\% or $\sim$70\% partial covering absorption to \KT~$\sim$12~keV plasma emission
by matter with a neutral hydrogen column density of $\sim$2$-$8$\times$10$^{21}$~\UNITNH,
while the spectrum outside of these dips is almost free of absorption.
This result suggests the presence of two distinct X-ray emitting spots in the \gcas\ system,
perhaps on a white dwarf companion with dipole mass accretion.
The partial covering absorbers may be blobs in the Be stellar wind, the Be disk, or rotating around the white dwarf companion.
Weak correlations of the softness ratios to the hard X-ray flux suggest the presence of stable plasmas at \KT~$\sim$0.9 and 5 keV,
which may originate from the Be or white dwarf winds.
The formation of a Be star and white dwarf binary system requires mass transfer between two stars; 
\gcas\ may have experienced such activity in the past.
\end{abstract}


\section{INTRODUCTION}

The classical Be stars are enigmatic massive stars with circumstellar disks,
whose formation is not yet understood \citep[e.g.][]{Rivinius2013}.
A subset of the class, the \gcas\ analogues, is particularly interesting with unusually hard (\KT~$\gtrsim$10~keV),
luminous (\LX~$\sim$10$^{32}-$10$^{33}$~\UNITLUMI) X-ray activity compared with that of normal main-sequence B stars \citep[see e.g.][]{SmithM2016}.
The origin of their high energy activity has been controversial for a long time.

As the name suggests, $\gamma$ Cassiopeia (\gcas) is the prototype of the \gcas\ class.
It is a B0.5~IVe star \citep{Tokovinin1997} with 
$R_{\ast} \sim$10~\UNITSOLARRADIUS\ and 
$L \sim$34000~\UNITSOLARLUMI\ \citep[e.g.,][]{Sigut2007}
at a distance of 188$\pm$4~pc \citep{vanLeeuwen2007}.
Doppler measurements of optical emission lines suggest that \gcas\ is a binary 
 with a nearly circular orbit ($e <$0.08) at a period of 203.5~days \citep{Nemravova2012,Smith2012a}.
The primary stellar mass is estimated from its effective temperature at $\approx$15~\UNITSOLARMASS \citep{Harmanec2000},
and the companion mass is estimated at $\sim$0.8$-$1~\UNITSOLARMASS\ from the orbital solutions,
assuming an orbital inclination of 45\DEGREE \citep{Nemravova2012,Smith2012a}.
The Be disk has been resolved with optical/near-infrared interferometry.
Its FWHM radius ranges between 1.5$-$7.25~$R_{\ast}$, depending on the measured wavelength
\citep[see references in][]{Rivinius2013}.
From elongation of the interferometric images, the disk inclination is estimated to be
$\approx$40$-$55\DEGREE \citep{Stee2012,Gies2007}.

Gamma Cas has been studied intensively in X-rays with almost all the available observatories 
\citep{White1982,Murakami1986,Parmar1993,Kubo1998,Smith1998,Owens1999,Robinson2000,SmithM2004,
Lopes_de_Oliveira2010,Smith2012a,Shrader2015,Motch2015}.
In all these observations the X-ray characteristics were found to be similar.
The light curve is highly variable, with a slowly varying ``basal" component on timescales of $\sim$10~ksec
and a rapidly varying ``shot" component on timescales of $\sim$10$-$100 sec.
Comparisons of the light curves in different energy bands have revealed strong variations in spectral hardness,
which are apparently uncorrelated to these flux variations.
The X-ray spectrum shows multiple emission lines, notably from hydrogen-like, helium-like and quasi-neutral $K$ shell iron.
Though these lines are apparently weak,
the broad-band spectrum does not show any non-thermal signature in continuum up to 100~keV
and can be formally reproduced with optically thin thermal emission from a \KT~$\sim$14~keV plasma
with \EM~$\approx$1$-$5$\times$10$^{55}$~\UNITEI\ and small metal abundance at $\sim$0.3~solar, suffering absorption at log \NH~$\sim$21$-$22~\UNITNH\ on average.
Nitrogen and oxygen $K$ shell lines and the iron $L$ shell line complex, resolved with high resolution grating spectra,
originates from a few relatively cool plasma components at \KT~$\sim$0.1$-$4~keV with \EM\ of 1$-$5$\times$10$^{54}$~\UNITEI.

Because of its high energy emission,
the presence of a degenerate companion, such as a neutron star or a white dwarf (WD), has been discussed.
The X-ray properties --- no signature of non-thermal emission, no pulsation, and moderate X-ray luminosity at $\sim$10$^{33}$~\UNITLUMI\
--- argue against the neutron star hypothesis.
The former two properties match with those of accreting WDs (cataclysmic variables),
but the X-ray luminosity is around the upper end of such systems 
and might be too high for mass accretion onto the known \gcas\ companion.
As \gcas\ has a circumstellar disk and shows apparently random, frequent X-ray shots,
an alternative mechanism of a magnetic dynamo driven by the star-disk differential rotation has also been proposed.
However, no strong magnetic field has been detected from \gcas,
nor has the validity of this mechanism been confirmed by theoretical investigations.
\citet{SmithM2016} discuss the pros and cons of these hypotheses in detail.

In this paper, we present our analysis of the color variation in the 0.3$-$10 keV X-ray band of \gcas\
observed with the \SUZAKU\ X-ray observatory \citep{Mitsuda2007} in 2011.
This observation was originally aimed at measuring the detailed spectral shape of the X-ray emission above 15~keV.
The initial result, combined with the archival \INTEGRAL\ data, which suggests no power-law component in the \gcas\ spectrum up to 100~keV,
is summarized in \citet{Shrader2015}.
The long-elapsed observation also enables us to monitor the spectral softness variation in the soft band in detail
and help discover that the variation, seen also in earlier observations,
is caused by rapid changes of the line of sight absorption to the hard, variable X-ray emission.

\section{Observation and Data Analysis}

\SUZAKU\ pointed at $\gamma$ Cas for $\sim$111~ksec starting on 2011 July 13, 00$^{\rm h}$ 06$^{\rm m}$ in UT.
During this observation, {\it Suzaku} used two instruments: the X-ray
Imaging Spectrometer \citep[XIS,][]{Koyama2007} at the focal plane of
the thin-foil X-Ray Telescope \citep[XRT,][]{Serlemitsos2007}, and the
Hard X-ray Detector \citep[HXD,][]{Takahashi2007,Kokubun2007}.  
In this paper, we focus on the XRT+XIS datasets.
The XIS consists of four X-ray CCD cameras, XIS0$-$3, three of which (XIS0, 2 and 3)
use front-illuminated (FI) CCD chips, while one (XIS1) uses a back-illuminated (BI) chip.
The FI chips have good hard X-ray sensitivity, covering the X-ray band between 0.5$-$10~keV,
while the BI chip has good soft X-ray sensitivity down to $\sim$0.3~keV.
During this observation, the sensors XIS0, 1 and 3 were working.
Gamma Cas was put at the XIS on-axis (XIS nominal) position
and the XIS net exposures are each 55.4~ksec.
The detailed observational set-up and the basic analysis procedure of this dataset 
can be found in \citet{Shrader2015}.
Throughout this paper, 
we use the analysis package HEASoft version 6.19 and Suzaku version 16 or later, and the CALDB versions, xis20110630 and xrt20110630,
for the data calibration.
We also use the {\tt Xanadu} package for the timing ({\tt xronos version 5.22}) and spectral \citep[{\tt xspec version 12.9.0i},][]{Arnaud1996}
analysis and assume the {\tt aspl} solar abundance table \citep{Asplund2009}\footnote{
We also tried the {\tt wilms} abundance table for the spectral fits (Section~\ref{subsec:phase_resolved_spectra}),
but these two models give negligible ($\lesssim$5\%) differences in the best-fit spectral parameters
except for the plasma elemental abundance, which is higher by $\sim$30\% with the {\tt wilms} abundance table.}.
In the timing analysis, the time origin is defined at Modified Julian Date 55755.0.

\section{Result}

\subsection{Light Curves and Softness Ratio Curves}

\begin{figure}[h]
\epsscale{0.9}
\plotone{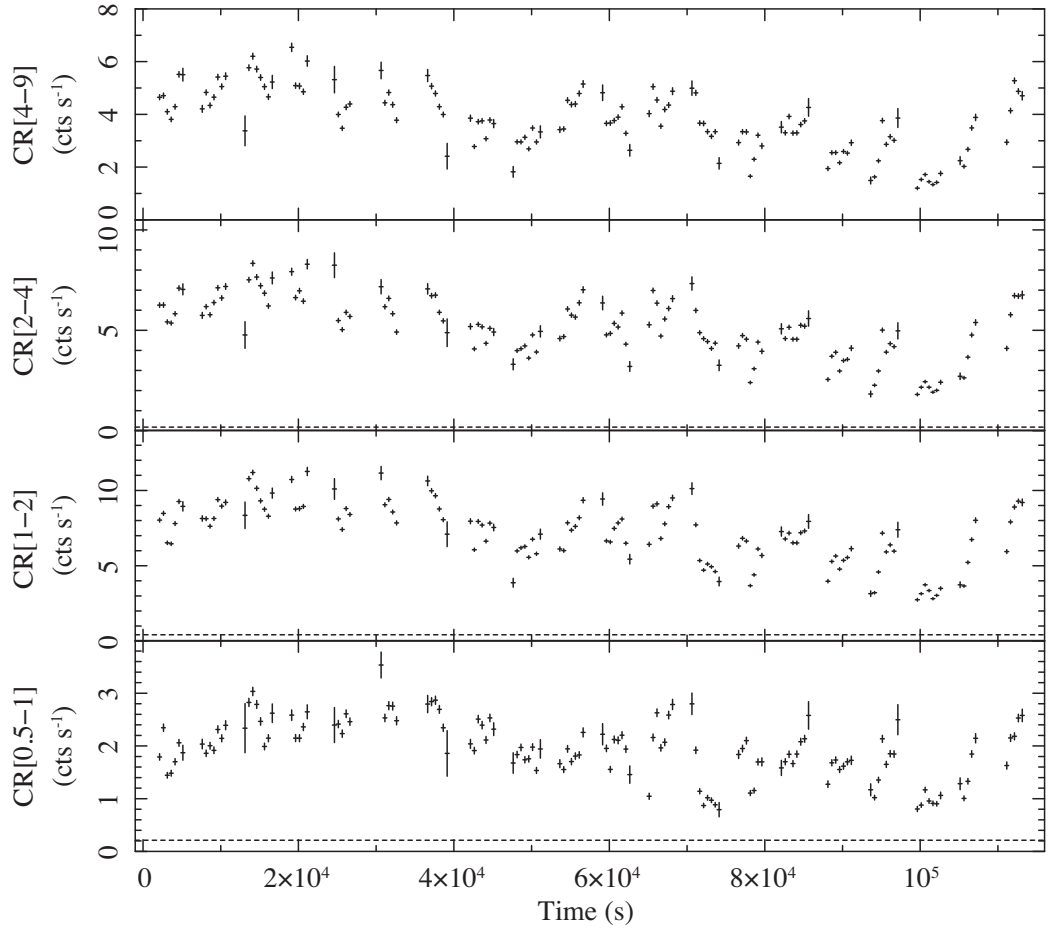}
\caption{XIS0+1+3 light curves of \gcas\ in the 4$-$9~keV ({\it top}), 2$-$4~keV ({\it 2nd}),  1$-$2~keV ({\it 3rd}),
and 0.5$-$1~keV ({\it bottom}) bands.
Each bin has 500~sec.
The time origin is at 55755.0~day in Modified Julian Date.
The dotted lines show the count rates of the stable emission estimated from the SR vs CR[4$-$9] plots
(Section~\ref{subsec:stable_component}, Figure~\ref{fig:color_flux_relation}).
\label{fig:light_curve}
}
\end{figure}

We generate XIS light curves of \gcas\ between 0.5$-$1~keV (count rate, CR[0.5$-$1]), 1$-$2~keV (CR[1$-$2]), 2$-$4~keV (CR[2$-$4])
and 4$-$9~keV (CR[4$-$9]) from the combined XIS (0, 1 and 3) data using the same procedure as described in Section~3.3 of 
\citet{Shrader2015} (Figure~\ref{fig:light_curve}).
All these light curves show intense fluctuations by $\approx$50\% on timescales of $\lesssim$1~ksec,
as also seen in earlier X-ray observations of \gcas\ \citep{White1982,Horaguchi1994,Kubo1998,Smith1998,Lopes_de_Oliveira2010,Smith2012a}.
During the observation, the fluxes in all energy bands
gradually decline to reach the lowest level at $\sim$100~ksec from the observation start, stay at that level for $\sim$5~ksec,
and then recover abruptly to the original flux level.
Such short, low flux intervals called ``cessation" were also seen in earlier observations \citep{Robinson2000,Robinson2002,SmithM2016}.

\begin{figure}[h]
\epsscale{1.0}
\plotone{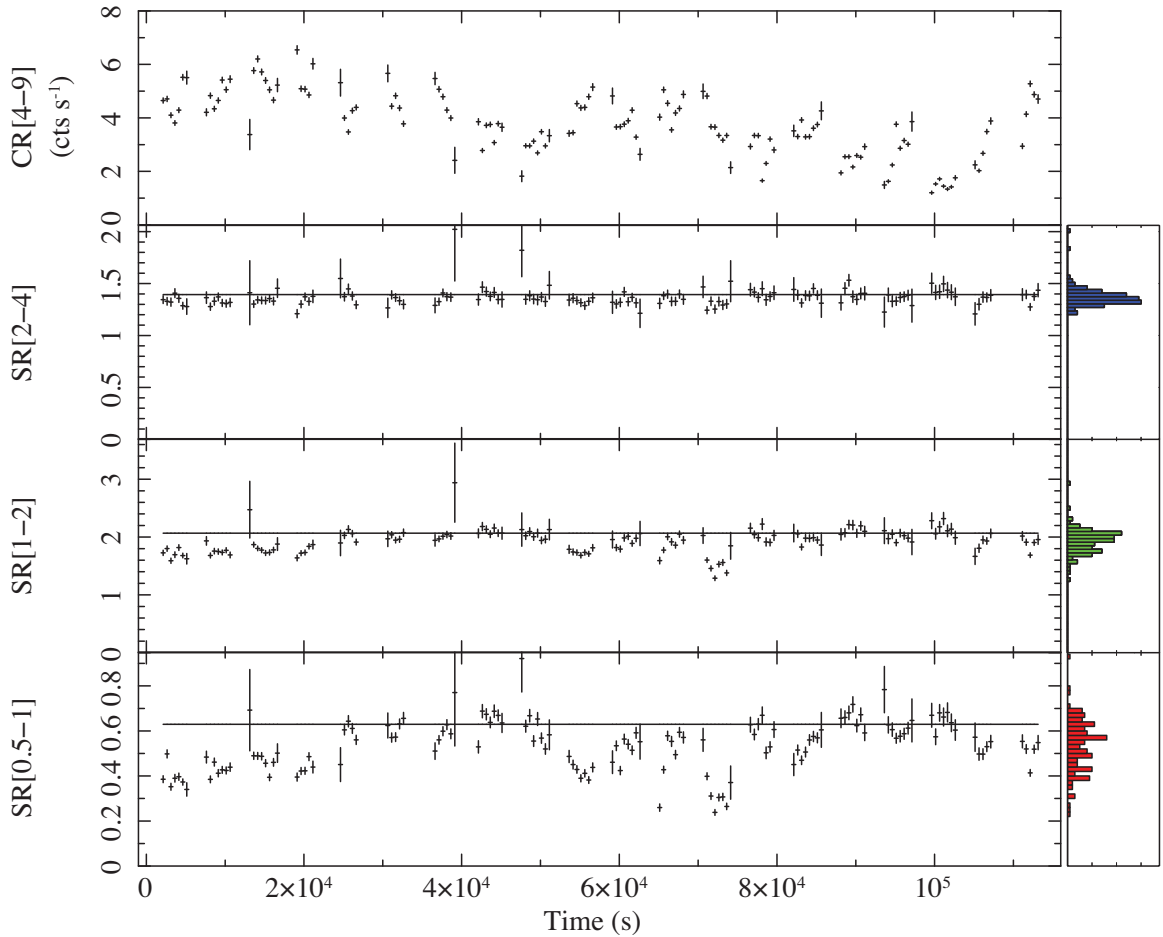}
\caption{XIS0+1+3 light curve in the 4$-$9~keV band ({\it top left}, the same as the {\it top} panel in Figure~\ref{fig:light_curve}),
time series of the softness ratios ({\it 2nd left}: SR[2$-$4], {\it 3rd left}: SR[1$-$2], {\it 4th left}: SR[0.5$-$1])
and distribution histograms of the softness ratios ({\it right}).
The horizontal bars in the 2nd, 3rd and 4th panels show average SRs during the baseline interval.
\label{fig:SR_histogram}
\label{fig:light_curve_ratio}
}
\end{figure}

In order to investigate the differences between these light curves, we produce softness ratio curves with respect to CR[4$-$9].
We define the softness ratios, SR[0.5$-$1], SR[1$-$2], and SR[2$-$4] as the divisions of CR[0.5$-$1], CR[1$-$2], and CR[2$-$4] by CR[4$-$9]
(i.e. SR[band] = CR[band]/CR[4$-$9] where ``band" is 0.5$-$1, 1$-$2, or 2$-$4).
The resulting color variability plots are shown in Figure~\ref{fig:light_curve_ratio}.
SR[2$-$4] is flat,
consistent with a constant model at above 97\% confidence (reduced $\chi^{2}$ = 1.23, 134 degrees of freedom, hereafter dof).
However, SR[1$-$2] begins to show dip-like structures; a constant model is inconsistent with this SR plot at a reduced $\chi^{2}$ of 10.1 (134 dof).
These dips are even more pronounced in the softest band SR[0.5$-$1], and
a constant model is rejected with a reduced $\chi^{2}$ of 25.0 (134 dof).
Figure~\ref{fig:light_curve_ratio} also demonstrates that, in the softer energy band,
these dips are deeper than the average SR value outside of these dips.
Similar spectral color variations seen as spikes in the hardness ratio plots (CR[4.5$-$10]/CR[0.3$-$1]) were
reported previously in \XMM\ observations \citep{Lopes_de_Oliveira2010,Smith2012a}.

The above result also implies that each SR variation has a baseline level at the highest clustering value ($\sim$0.6 for SR[0.5$-$1] and $\sim$2.1 for SR[1$-$2]).
There are important reasons to choose this level as the baseline instead of a lower one.
First, it constitutes a clear peak at $\sim$2.1 for SR[1$-$2] and a maximum at $\sim$0.6 for SR[0.5$-$1] (see histograms at the right panels of Figure~\ref{fig:light_curve_ratio}).
Second, all significant transitions of the SR value start or end at these baseline levels.
For example, if we choose the baseline level at $\sim$0.4 in SR[0.5$-$1],
transitions at $\sim$65 and 73~ksec just pass this level.
This level does not seem to be important physically and/or phenomenologically.
Third, our baseline levels are consistent with that of zero absorption to \KT~$\sim$12~keV X-ray emission,
which is the major X-ray emission component of \gcas\ (see section~\ref{subsec:phase_resolved_spectra}).
We therefore believe our choice of the baseline levels is the most physically reasonable.

Such spectral hardening could be produced by plasma heating,
but these softness dip events do not occur coincidentally with intrinsic flux increases traced with CR[4$-$9],
which would indicate plasma heating (Figure~\ref{fig:light_curve_ratio}).
They can be, instead, produced by changes in the absorption column.
The photo-electric absorption cross section depends on the X-ray photon energy as $E^{-3}$ \citep{Morrison1983}.
Cold matter in the line of sight with columns of \NH\ $\sim$10$^{21}-$10$^{22}$~\UNITNH\ 
absorb only soft X-rays below $\sim$2~keV.

\begin{figure}[h]
\epsscale{1.0}
\plotone{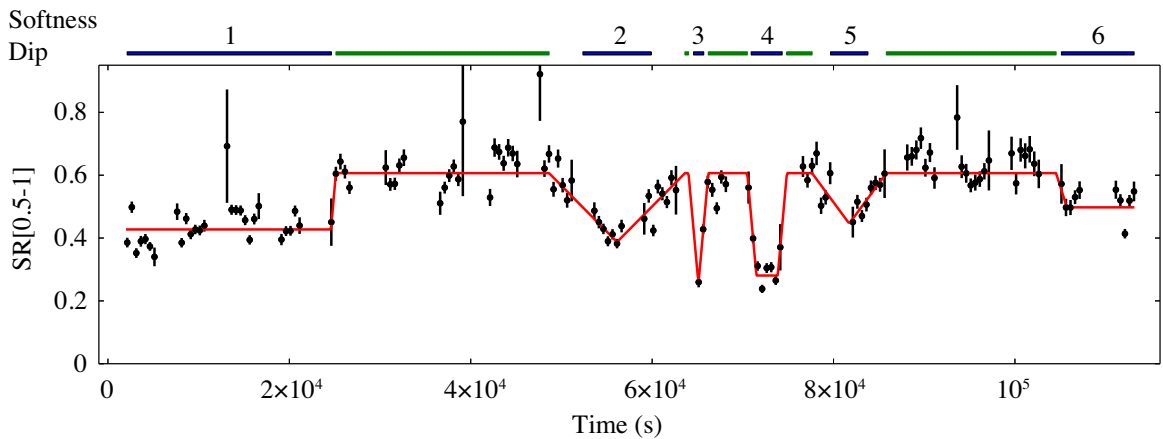}
\caption{SR[0.5$-$1] fitted by the empirical softness dip model ({\it solid red line}).
Each softness dip interval employed for the spectral analysis is shown as a horizontal bar, while
the baseline interval is shown as green bars.
\label{fig:SR_curve_fit}
}
\end{figure}

In SR[0.5$-$1], we identify six clear softness dips, two of which (dips 1 and 6) overlap with the start or end of the observation.
Softness dips 1, 3, 4 and 6 show rapid fall and rise, while the others have relatively slow transitions.
Softness dips 1, 4 and 6 have flat bottoms, while the others instantly transition from fall to rise.
Though parts of these dips are missed because of the gaps in the data, their shapes appear to be symmetric.
We therefore construct a phenomenological model, describing each softness dip by a symmetrical trapezoid with four parameters --- 
center time ($t_{\rm c}$), flat bottom duration ($T_{\rm fb}$), fall/rise transition duration ($T_{\rm tr}$),
and depth ($SR_{\rm d}$).
Then, the shape of the softness dip $i$ can be formulated as:
\begin{alignat}{2}
SR^{i}(t) &= 0 && (T_{\rm fb}^{i}/2 + T_{\rm tr}^{i} < \left|t-t_{\rm c}^{i}\right|)\\
		    &= -\frac{(T_{\rm tr}^{i}+T_{\rm fb}^{i}/2-\left|t-t_{\rm c}^{i}\right|)}{T_{\rm tr}^{i}}SR_{\rm d}^{i}~~~~&&(T_{\rm fb}^{i}/2 < \left|t-t_{\rm c}^{i}\right| \leq T_{\rm fb}^{i}/2 + T_{\rm tr}^{i})\\
		    &= -SR_{\rm d}^{i} &&(\left|t-t_{\rm c}^{i}\right| \leq T_{\rm fb}^{i}/2)
\end{alignat}
SR[0.5$-$1] can be described as the sum of the six softness dips on top of the baseline level ($SR_{\rm bl}$) as follows.
\begin{eqnarray}	    
SR(t) &= SR_{\rm bl} + \sum\limits_{i = 1}^{6} SR^{i}(t)
\end{eqnarray}
We fit this model to SR[0.5$-$1].
The best-fit result is shown in Figure~\ref{fig:SR_curve_fit} and Table~\ref{tbl:SR_curve_fit}.
$T_{\rm tr}$ is roughly divided into short (0.4$-$1~ksec) and long (4$-$7~ksec) timescales.
$T_{\rm fb}$ is distributed in a broad range from zero to $>$20~ksec.

\subsection{Stable Spectral Component}
\label{subsec:stable_component}
\begin{figure}[h]
\epsscale{1.0}
\plotone{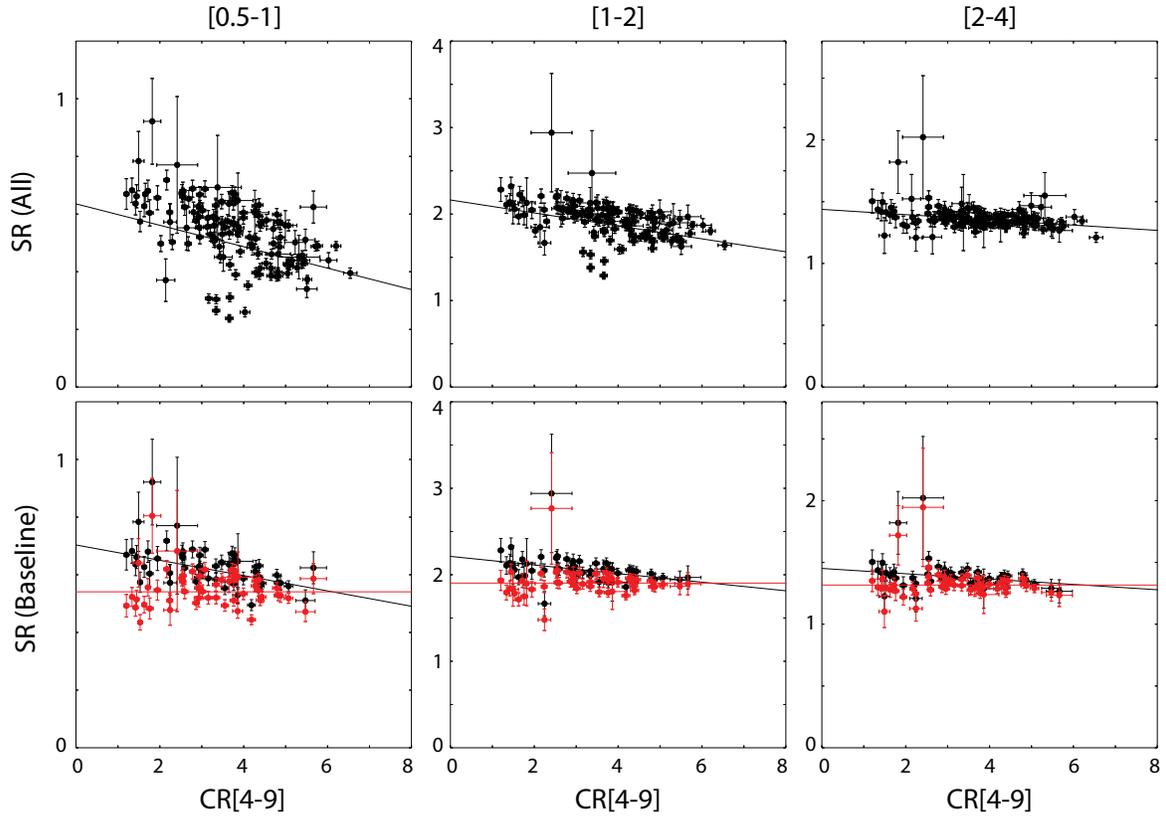}
\caption{The softness ratios ({\it left}: SR[0.5$-$1], {\it middle}: SR[1$-$2], {\it right}: SR[2$-$4]) against the 4$-$9~keV flux (CR[4$-$9])
for all data points ({\it top}) and data points obtained during the baseline interval ({\it bottom}).
The {\it bottom} panels show both the original SRs ({\it black}) and the SRs after the soft CR reduction thought to be due to the soft, stable component ({\it red}).
Each solid line shows the best linear fit to the corresponding data points.
\label{fig:color_flux_relation}
}
\end{figure}

We further exploit the information in the SRs and
find a weak but significant correlation between each SR and CR[4$-$9];
at each energy band, the SR is higher in lower CR[4$-$9] (Figure~\ref{fig:color_flux_relation}, upper panels).
The same plot but only with data points outside of the softness dips (hereafter called the baseline interval:
$\sum[t_{\rm c}^{i} + (T_{\rm fb}^{i}/2 + T_{\rm tr}^{i}), t_{\rm c}^{i+1} - (T_{\rm fb}^{i+1}/2 + T_{\rm tr}^{i+1})]$)
shows a linear correlation more clearly (see the black data points in the bottom panels of Figure~\ref{fig:color_flux_relation}).
In fact, a linear fit to the SR[0.5$-$1] vs.\ CR[4$-$9] plot improves from 22.3 (dof: 133) to 2.45 (dof: 52) in reduced $\chi^2$ value
by excluding the softness dip data points.
Therefore, this correlation probably is not produced by coincidental dip events.
Such correlations can be produced if the plasma temperature tends to be higher or the absorption tends to be stronger at higher CR[4$-$9].
However, a similar plot for SR[15$-$40] in Figure~3 of \citet{Shrader2015} did not show any dependence on CR[4$-$9],
even though the temperature increase should be most clearly seen in the highest energy range.
It is also difficult to imagine a straightforward physical mechanism that could produce stronger absorption in the higher CR[4$-$9] range.

We suggest here that this correlation is most consistently explained by soft, stable X-ray emission.
The presence of relatively cool plasmas at \KT $\sim$0.1$-$4~keV is suggested from nitrogen $K$, oxygen $K$, and iron $L$ emission lines
detected in high resolution grating spectra \citep{SmithM2004,Smith2012a}.
If some or all of these components are present stably during the \SUZAKU\ observation,
their emission should produce the observed SR and CR[4$-$9] relation.
We derive the contribution of these stable components, assuming that the hard, variable component does not change the spectral shape with CR[4$-$9], 
based on no apparent correlation of SR[15$-$40] to CR[4$-$9] \citep[Figure~3 of ][]{Shrader2015}.
We reduce an equal amount of count rates from CR in each band below 4~keV
until the linear fit of each SR vs.\ CR[4$-$9] plot is flat.
The result is shown as the red points in the bottom panels of Figure~\ref{fig:color_flux_relation}, obtained by reducing the soft CR by
0.213 (SR[0.5$-$1]), 0.418 (SR[1$-$2]) and 0.184 (SR[2$-$4]) \UNITCPS.

\begin{figure}[h]
\epsscale{1.0}
\plotone{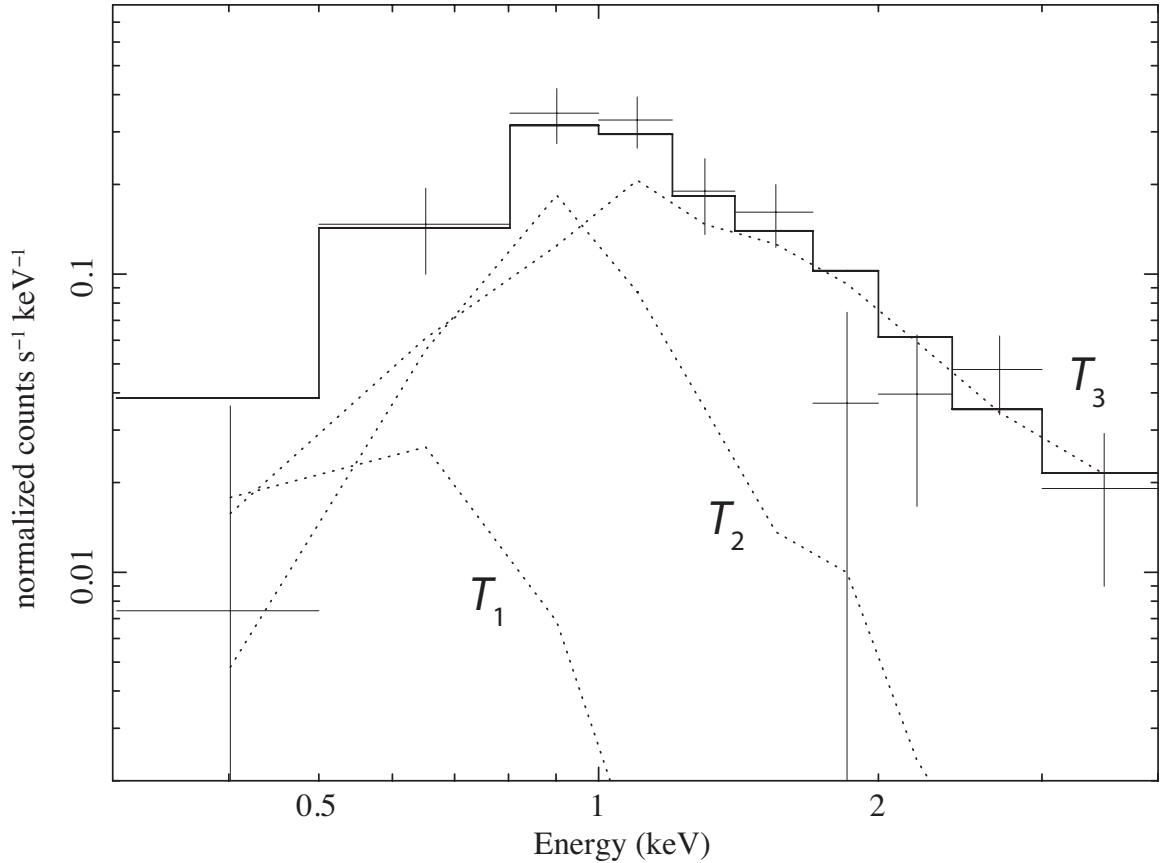}
\caption{Stable emission spectrum for the XIS1, estimated from the SR vs.\ CR[4$-$9] plots.
The solid line shows the sum of the $T_{1}$, $T_{2}$ and $T_{3}$ components of the best-fit OBS1 spectrum in \citet{Smith2012a}, 
normalized by 0.64 to fit the \SUZAKU\ spectrum.
The dotted lines show the normalized individual components.
\label{fig:spec_stable}
}
\end{figure}

Since this approach works apparently well,
we re-define 10 energy bands between 0.3$-$4 keV
and measure the reduction amount for each bin in the same way.
We use all the XIS (0,1,3) data between 0.5$-$4~keV
and derive the reduction amount for the XIS1 from the count rate ratio.
The FI sensors are not sensitive below 0.5~keV, so we only use XIS1 for the 0.3$-$0.5 keV bin.
Their 1$\sigma$ errors are evaluated from the cases when a linear fit to the plot includes a flat solution within a 1$\sigma$ error.
Figure~\ref{fig:spec_stable} shows the stable emission spectrum for XIS1 estimated from this method.

The stable spectrum peaks at $\sim$0.9~keV and extends up to $\sim$4 keV.
If this spectrum originates from optically thin thermal plasma emission,
these profiles require, at least two temperature components at \KT~$\sim$0.9~keV and $\gtrsim$3 keV.
These temperatures are similar to the $T_{2}$ (\KT~$\sim$0.6~keV) and $T_{3}$ (\KT~$\sim$4~keV) components 
in the \XMM\ observations in 2010 \citep{Smith2012a}.
In fact, the sum of the three cooler components in the best-fit {\it XMM}/OBS1 spectrum ($T_{1}$: \KT~$\sim$0.11~keV, $T_{2}$ and
$T_{3}$)\footnote{The $EM_{2}$ value is reduced by an order of magnitude from the value in Table 4 of \citet{Smith2012a}, 
to correct the typological error of the exponent of the unit (R. Lopes de Oliveira, private comm.).}
fits nicely to this stable spectrum by multiplying the model normalization by 0.64 (Figure~\ref{fig:spec_stable}).
This result suggests that the stable spectrum has the same origin as the $T_{2}$ and $T_{3}$ components.
The difference in normalization may be due to time variation or the different measurement method ---
we estimate the stable spectrum from the broad-band color variation, 
while \citet{Smith2012a} estimated these cool components from emission lines in the time averaged high resolution spectra.
The minor $T_{1}$ component might also be a part of this stable emission, but we cannot judge it conclusively from the \SUZAKU\ data.

\subsection{Phase Resolved Spectra}
\label{subsec:phase_resolved_spectra}

\begin{figure}[h]
\epsscale{1.0}
\plottwo{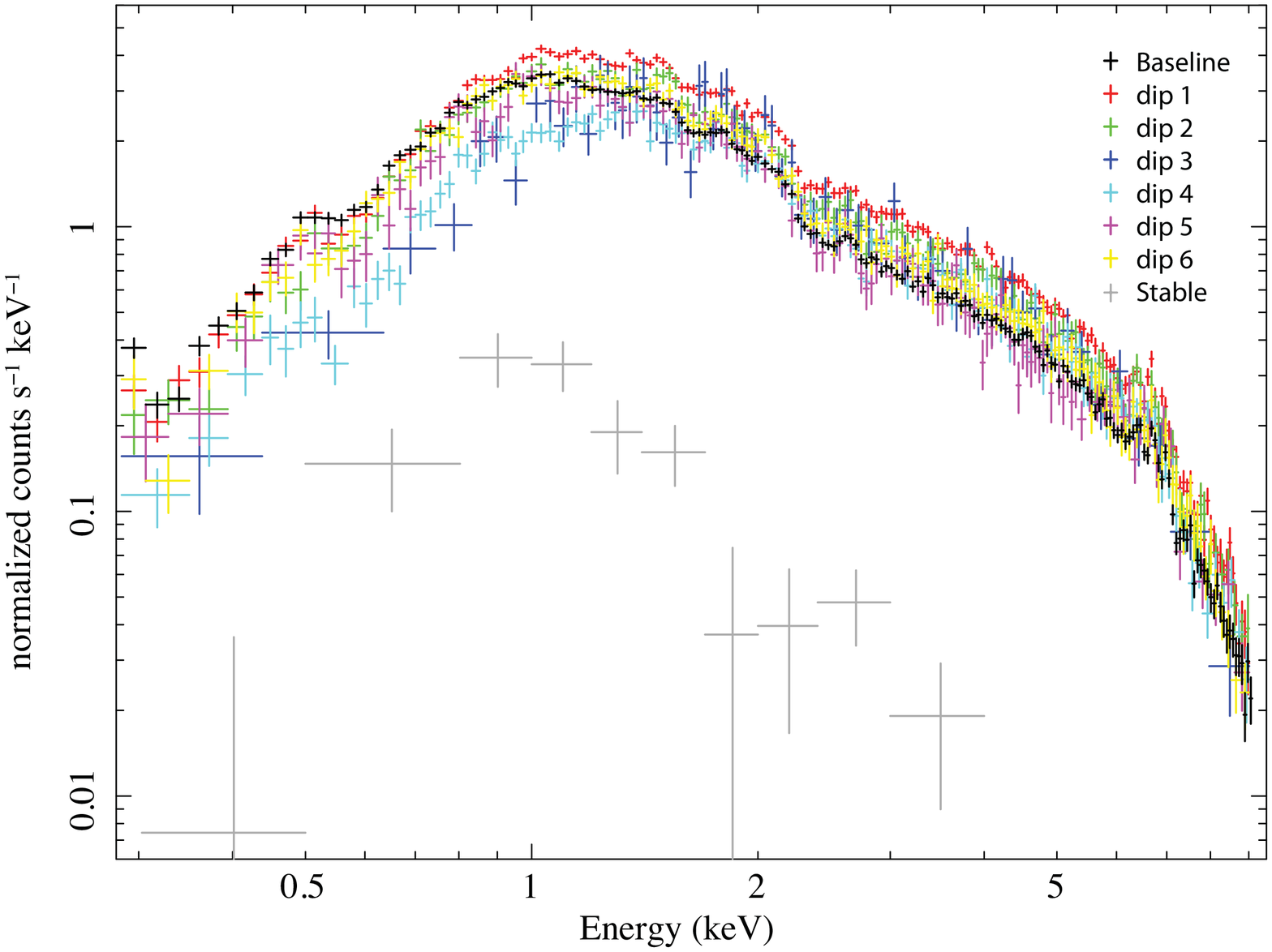}{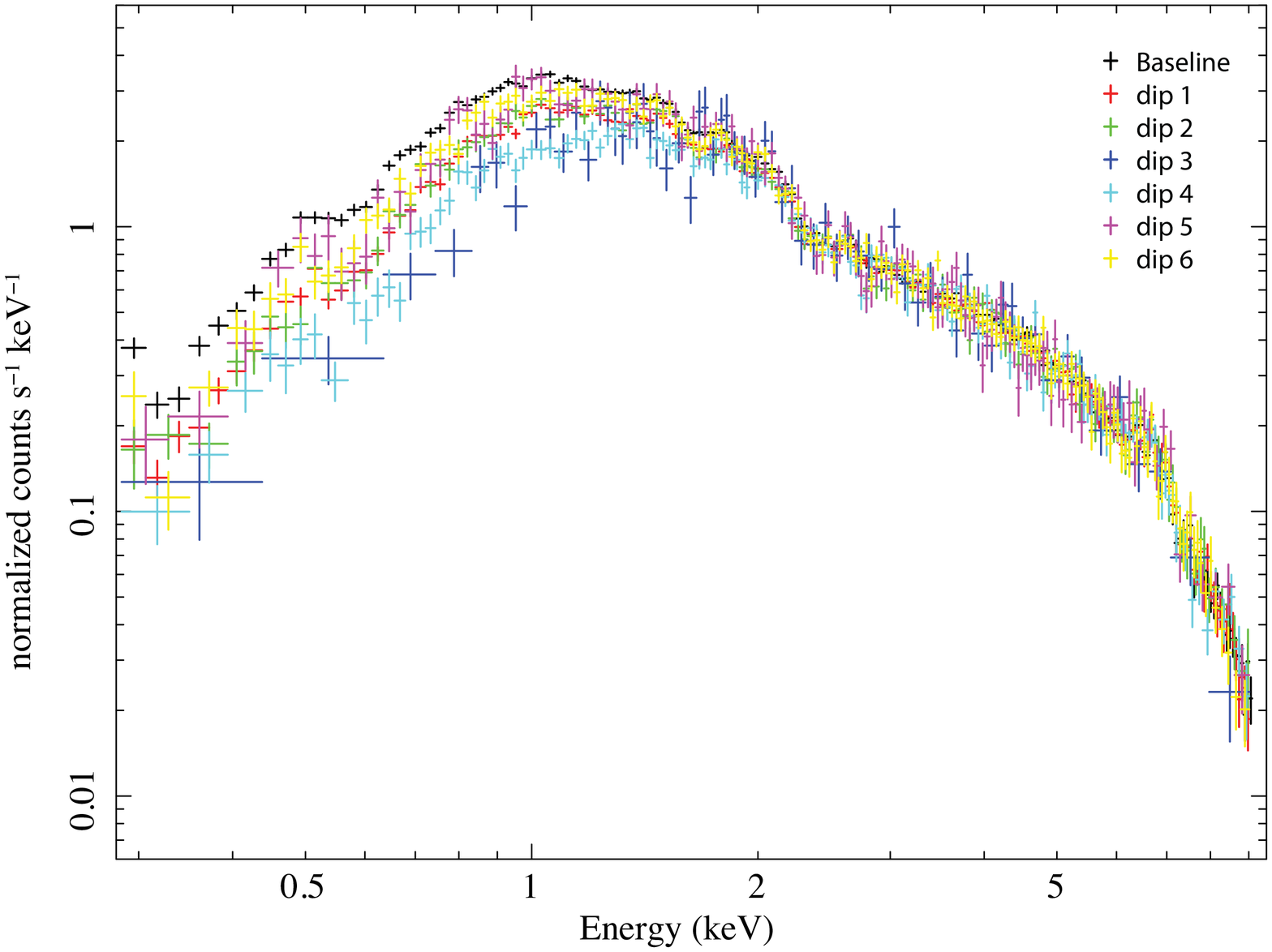}
\caption{{\it Left}: XIS1 spectra of the baseline interval, all of the softness dips and the stable emission. 
{\it Right}: The same spectra, normalized with the best-fit normalization values of the hard variable component.
\label{fig:spec_dip}
}
\end{figure}

To understand the nature of the softness dips,
we extract XIS spectra during each dip, which is defined as the interval
between the middle points of the fall and rise in the best-fit SR[0.5$-$1] model
(i.e., $[t_{\rm c}^{i} - (T_{\rm fb}^{i} + T_{\rm tr}^{i})/2, t_{\rm c}^{i} + (T_{\rm fb}^{i} + T_{\rm tr}^{i})/2]$).
We also extract XIS spectra of the baseline interval.
The FI (XIS0 and 3) spectra of each interval are combined.
The XIS1 spectra of the baseline interval, all the softness dips, and the stable emission are overlaid in the left panel of Figure~\ref{fig:spec_dip}.
The baseline and softness dip spectra have a factor of $\lesssim$2 flux variations in both the soft and hard bands.
In the right panel of Figure~\ref{fig:spec_dip}, we show the XIS1 spectra of the baseline interval and the softness dips, normalized with the normalization parameter 
of the variable thermal component to the baseline spectrum.
The softness dip spectra clearly have stronger soft-band cut-offs than the baseline spectrum,
but they do not show significant changes in the hard band slope or the flux ratio of the hydrogen-like and helium-like iron $K$ emission lines.
This plot further supports our conclusion that the softness dips are produced by changes in the absorption, and not the plasma temperature.

We then fit all these spectra simultaneously.
For the stable emission, we assume an absorbed two-temperature thermal model --- 
({\tt apec} + {\tt apec}){\tt TBabs}.
For the hard variable emission, we apply a one-temperature thermal ({\tt apec}) model 
plus two Gaussian lines at 6.4~keV and 7.04 keV for iron $K_{\alpha}$ and $K_{\beta}$ fluorescence,
suffering common absorption.\footnote{\citet{Smith2012a} suggested the presence of an extremely embedded hard thermal component.
We consider it as Compton reflection of the hard variable thermal emission, which should be accompanied by fluorescent iron emission lines.
Our preliminary fit to the \SUZAKU\ spectra suggest negligible contribution of this component, possibly due to different geometry of the X-ray 
emitting source and the cold matter or reflector \citep{Morihana2014}.
We, therefore, do not include this component in our model.}
We tie all the thermal emission parameters except for the normalizations.
We vary the normalization of the fluorescent iron $K_{\alpha}$ line
and tie the normalization of the fluorescent iron $K_{\beta}$ line to 12\% of this value.

\begin{figure}[h]
\epsscale{1.0}
\plotone{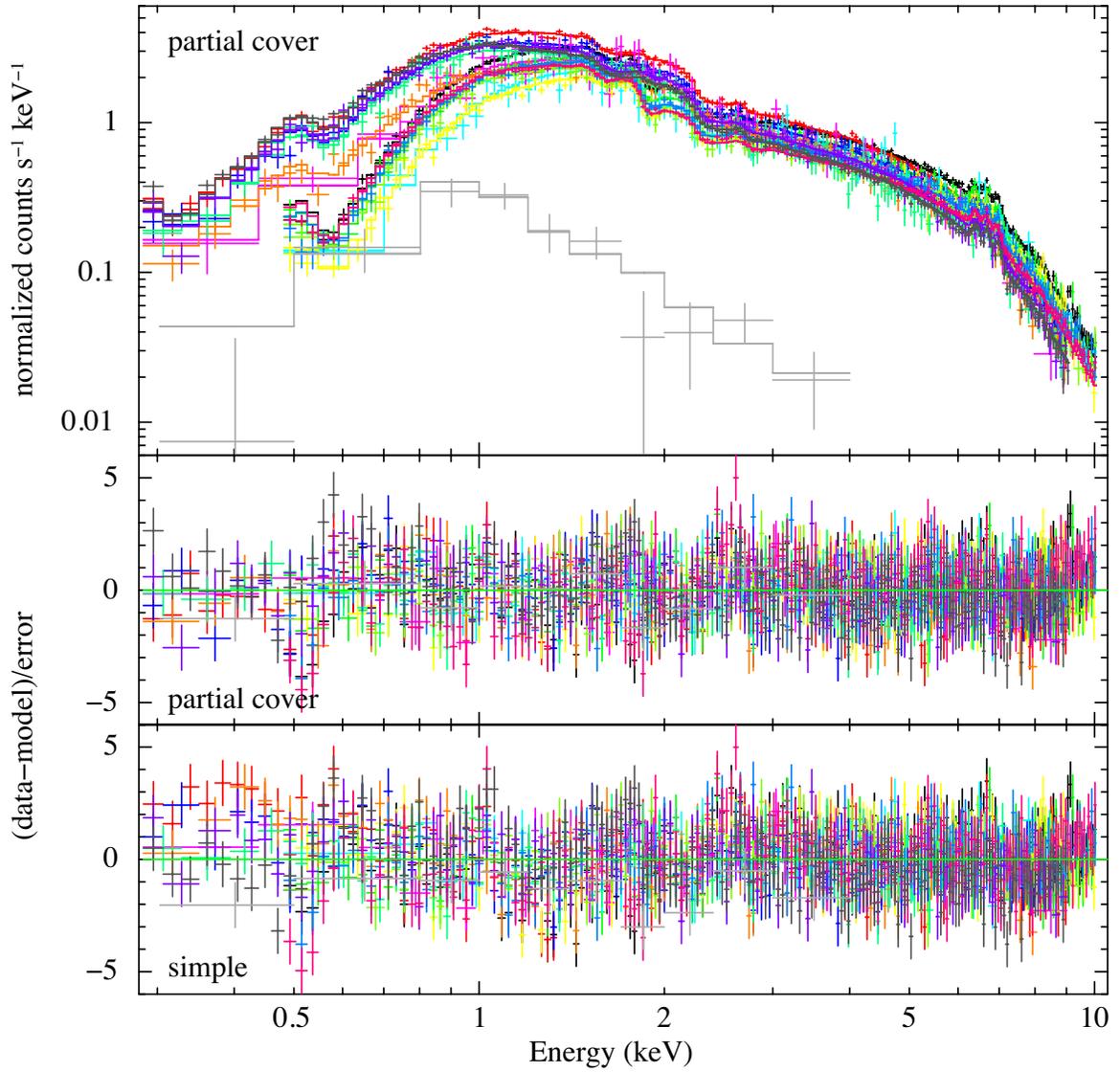}
\caption{FI (XIS0+3) and XIS1 spectra of the baseline interval, all of the softness dips and the stable emission along with their 
best-fit model in the solid lines, assuming partial covering absorbers ({\it top}).
The figure also shows the residuals to the best-fit models assuming the partial covering absorbers 
({\it middle}) and simple absorbers ({\it bottom}).
The model with the partial covering absorbers has smaller residuals below $\sim$0.8~keV.
\label{fig:spec_model_fit}
}
\end{figure}

We first try simple absorption \citep[{\tt TBabs},][]{Wilms2000} for the variable component.
The spectra above 0.8~keV fit well, but 
the softness dip spectra below 0.8~keV show significant deviations with a bad chi-square value 
(reduced $\chi^{2}$ = 1.50, dof = 2214, see the residual in the bottom panel of Figure~\ref{fig:spec_model_fit}).
This looks as if less absorbed emission emerges during each softness dip interval.
Since the stable emission is constrained,
this emission should be a part of the variable component, which is not absorbed during the softness dip event, so the absorption is partial.
We therefore apply a partial covering model, {\tt pcfabs}, for each softness dip,
in addition to simple absorption common for all spectra, which is required for absorption to the baseline spectra.
Then, the deviation decreases substantially
(reduced $\chi^{2}$ =1.35, d.o.f = 2208,
Table~\ref{tbl:dip_spec_pcf} and the top and middle panels of Figure~\ref{fig:spec_model_fit}).
In summary, the applied model to each spectrum $F_{{\rm X}, i}$ ($i$: baseline, 1$-$6) is:
\begin{alignat}{2}
F_{{\rm X}, i} &= F_{\rm X}[V_i] + F_{\rm X}[S]\\
F_{\rm X}[V_i] &= ({\tt apec}\{kT[V], Z[V], EM[V_i]\} + {\tt gauss}_{6.40}\{F_{{\rm X, FeI K\alpha}}[V_i]\} + {\tt gauss}_{7.04}\{0.12 F_{{\rm X, FeI K\alpha}}[V_i]\}) \nonumber \\
& ~~~~~~~~  {\tt pcfabs}\{N_{\rm H, PC}[V_i], CF_{\rm PC}[V_i]\}*{\tt TBabs}\{N_{\rm H}[V]\} \\
F_{\rm X}[S] &= ({\tt apec}\{kT[S_{\rm C}], EM[S_{\rm C}]\} + {\tt apec}\{kT[S_{\rm H}], EM[S_{\rm H}]\}){\tt TBabs}\{N_{\rm H}[S]\}
\end{alignat}
where square brackets show the emission component --- $V_{(i)}$: ($i$-th) variable component, $S_{\rm (C/H)}$: (cool/hot) stable component ---
and curly brackets show free parameters for each spectral component.
The spectral parameters with the subscript $i$ are fitted independently between the spectra, while the others are tied between them.
The best-fit plasma temperature of the variable component is \KT~$\sim$12.2~keV, typical of \gcas.
The absorption columns of the partial absorbers are between 2.4$-$8.1$\times$10$^{21}$~\UNITNH,
while the covering fractions are localized in two ranges, $\sim$40\% and $\sim$70\%.
The common absorption, representing the \NH\ to the baseline spectra, is very low at $\sim$3.8$\times$10$^{20}$~\UNITNH\
and close to the column density to the Be star measured from the UV spectroscopy 
\citep[1.45$\pm$0.3$\times$10$^{20}$~\UNITNH,][]{Jenkins2009}.
There is little local absorption outside of the softness dip intervals.
The equivalent width of the fluorescent iron $K_{\alpha}$ line does not vary significantly at $\sim$60~eV,
suggesting that the softness dip absorbers are not responsible for the iron fluorescence emission.

\subsection{Pulse Search for Data above 2 keV}
Earlier pulsation analysis with the \XMM\ data was performed for the broad 0.8$-$10~keV band \citep{Lopes_de_Oliveira2010,Smith2012a}.
Since soft X-ray emission from \gcas\ is affected by the dip events,
we undertake a new pulse search only above 2 keV using the standard Fourier analysis.
However, we find no significant pulse in our analysis, either.

\section{Discussion}

We find that hard, variable X-ray emission from \gcas\ is frequently occulted by absorbers 
with log \NH~$\sim$21$-$22~\UNITNH\ on timescales of a kilo-second or longer during the \SUZAKU\ observation.
Earlier \XMM\ observations of \gcas\ also reported similar variations in the hardness ratio \citep{Lopes_de_Oliveira2010,Smith2012a}.
Gamma Cas and HD~110432, a \gcas\ analogue, show changes in time-averaged X-ray absorption between observations \citep[e.g., Figure~3 in][]{SmithM2016},
which are perhaps caused by variations in numbers and column densities of softness dip absorbers for each observation.
Rapidly changing X-ray absorbers are probably a common feature in \gcas\ analogues.

All softness dips are symmetrical.
This can be explained if, in each softness dip event, a blob-like absorber passes over a discrete X-ray emitting region with a constant projection velocity.
The triangular dips 2, 3 and 5 can be produced by absorbers that block only a part of an X-ray emitting region.
The trapezoidal dips 1, 4 and 6 can be produced in the following three cases.
In the first case, an absorber moves over an X-ray emitting region in such a way that its covering factor is constant during the flat bottom,
but this situation seems unlikely to occur.
In the second case, an absorber is fully inside of a larger X-ray emitting region on projection during the flat bottom.
In this case, $T_{\rm tr}$ relates to the projection size of the absorber, 
while $T_{\rm fb}$ corresponds to the travel length of the absorber over the X-ray emitting region,
i.e., the minimum size of the X-ray emitting region.
The size ratios of the X-ray emitting region over the absorber ($T_{\rm fb}$/$T_{\rm tr}$) are
$\gtrsim$45, 2.4, 7 for softness dips 1, 4 and 6.
Assuming two-dimensional symmetry on the projection surface,
the areal ratios are $\gtrsim$2000, 5, and 50.
In this case, the covering factors of the partial covering absorbers are $\lesssim$0.05, 17 and 1.8\%, 
which contradict those measured from the spectral fits, $\sim$41, 65, and 35\%.
In the third case, an absorber is larger than an X-ray emitting region 
and fully obscures it during the flat bottom.
In this case, $T_{\rm tr}$ relates to the projection size of the X-ray emitting region,
and $T_{\rm fb}$ corresponds to the absorber size.
The covering factor of the absorber to this X-ray emitting region becomes unity, so
there should be other uncovered discrete X-ray emitting regions that produce the flat-bottom emission.
We conclude that the third case --- multiple X-ray emitting regions, one of which is fully covered during each flat bottom
--- should be the only viable solution.
Since the covering factor of each dip is $\gtrsim$40\%,
the covered X-ray emitting region should be a major X-ray spot.

Based on the above consideration,
the length of the X-ray emitting region ($L_{\rm proj}$) is expressed in terms of the projected velocity of the absorber ($v_{\rm proj}$) 
and the transition timescale as $L_{\rm proj} \sim v_{\rm proj} T_{\rm tr}$.
The SR[0.5$-$1] fit suggests at least two distinct groups in $T_{\rm tr}$ at 0.4$-$1~ksec and 4$-$7~ksec.
This might suggest two kinds of absorbers with different projection velocities or two different X-ray emitting regions occulted by the absorbers.
On the other hand,
the absorber's projection length ($l_{\rm proj}$) and the line of sight depth ($l_{\rm d}$) 
are related to the dip parameters $T_{\rm fb}$ and \NH\ as
$l_{\rm proj} \sim v_{\rm proj} T_{\rm fb} \sim (T_{\rm fb}/T_{\rm tr})L_{\rm proj}$ and 
${l_{\rm d}} \sim$ \NH$/\rho_{\rm abs}$ where $\rho_{abs}$ is the absorber's density.
Their ratio is $l_{\rm proj}/{l_{\rm d}} \sim$ $\rho_{\rm abs}v_{\rm proj}$ $T_{\rm fb}$/\NH $\sim$
$\rho_{\rm abs}L_{\rm proj}T_{\rm fb}/(T_{\rm tr}$\NH).
The value $T_{\rm fb}/(T_{\rm tr}$\NH) ranges by a factor of $\gtrsim$25 for dips 1, 4 and 6.
If $\rho_{\rm abs}$ and $L_{\rm proj}$ do not change significantly between the absorbers, 
the absorber may not be three dimensionally symmetric, i.e. $l_{\rm proj} \nsim {l_{\rm d}}$.

\begin{figure}[h]
\epsscale{1.0}
\plotone{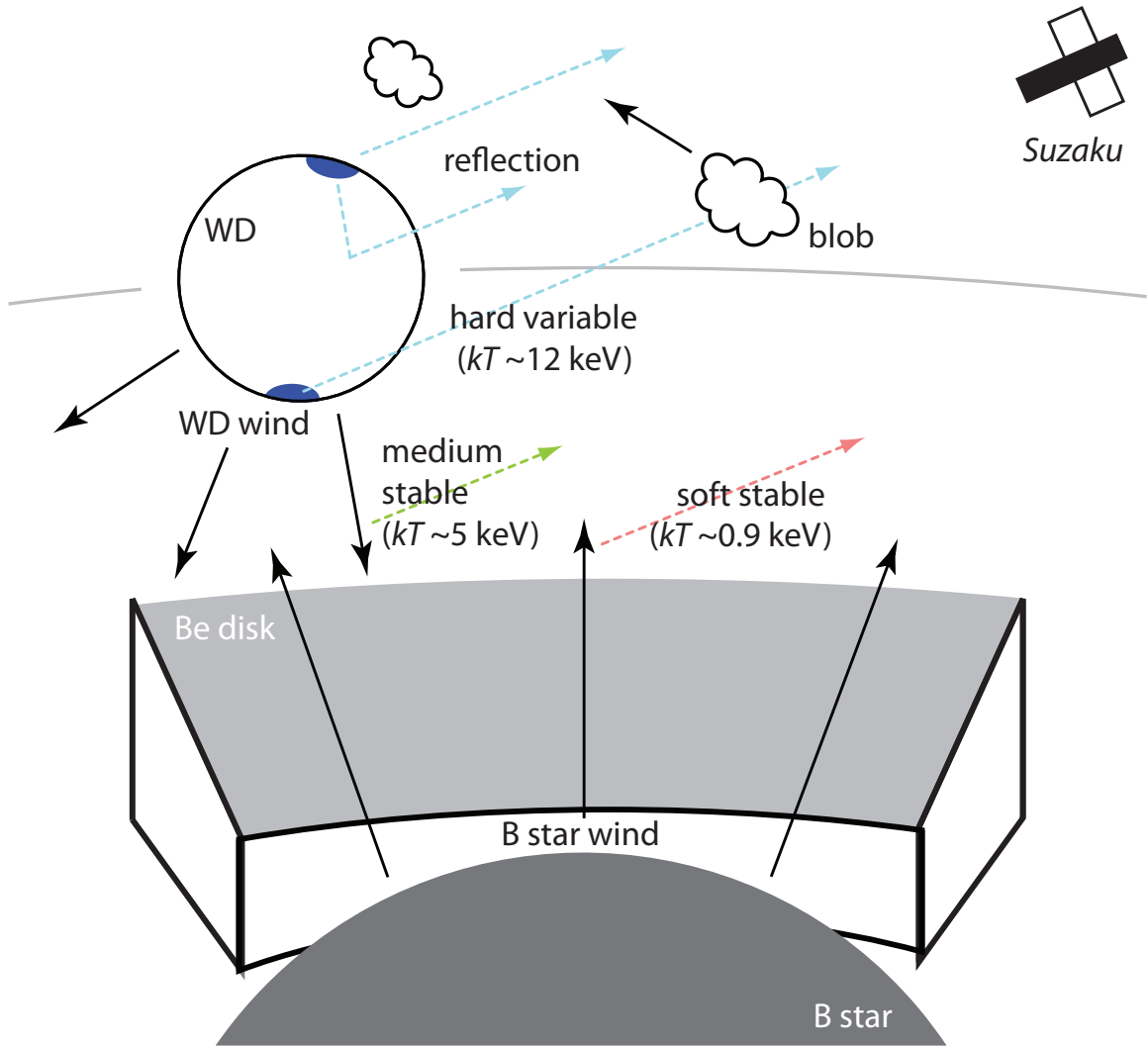}
\caption{Possible geometry of the X-ray emitting and absorbing components.
\label{fig:geometry}
}
\end{figure}

The best-fit result of the partial covering model suggests two covering factors, $\sim$0.4 and $\sim$0.7.
This result may not be consistent with the magnetic star-disk dynamo theory, in which X-rays originate from multiple magnetic loops 
on the Be stellar surface \citep[see][]{SmithM2016}.
Assuming that the magnetic loops spread evenly over the Be star surface,
a single X-ray absorber should have a size of $\sim$0.4 stellar radius ($\sim$4~\UNITSOLARRADIUS)
and move with a projection velocity of up to $\sim$5000~\UNITVEL,
which are unrealistic for the wind or disk of the Be star.
On the other hand, this result seems to fit with the mass-accreting compact star,
which has small numbers of discrete X-ray emitting regions.
Interestingly, the sum of the two covering factors is close to unity, perhaps suggesting that
one of two distinct X-ray emitting regions on the dipole magnetic poles is occulted occasionally.
The cessation seen at $\sim$10$^{5}$~sec is also similar to the flux drops observed in 
the high mass X-ray binary, Vela X-1, which can be explained from the propeller effect --- inhibition of 
mass accretion due to the change of the ram pressure of the infalling gas \citep{Kreykenbohm2008}.
As described in the introduction,
the companion is perhaps not an accreting neutron star since it shows no evidence of either a power-law tail up to 100 keV or pulsations
\citep[][this work]{Lopes_de_Oliveira2010,Smith2012a,Shrader2015},
although a recent model argues the existence of neutron stars with such spectral characteristics like those of \gcas\ (Postnov K. 2016, in preparation).
A \KT~$\sim$12~keV thermal plasma cannot be produced by free-fall accretion onto a sub-dwarf with log $g\lesssim$6.5 \citep{Heber2009}.
We, hence, propose that the hard, variable X-ray emission originates from accretion hot spots of a $\sim$1~\UNITSOLARMASS\ WD companion (Figure~\ref{fig:geometry}).

The known WD binary systems that fits most of the above characteristics are the intermediate polars (IP);
the IPs are composed of a late-type donor star and a weakly magnetized accreting WD ($B <$10~MG)
and their X-ray luminosities range up to $\sim$10$^{33}$~\UNITLUMI \citep{Yuasa2010}.
However, they tend to have higher plasma temperatures (\KT~$>$20~keV) and larger absorption columns 
(log \NH $\sim$22$-$23~\UNITNH) than \gcas.
An exception is EX Hydrae with low \KT\ ($\sim$13~keV) and \NH\ ($\sim$7$\times$10$^{21}$~\UNITNH),
which may be realized with mass accretion on relatively large areas and
accretion shocks at a high altitude \citep[$\approx$0.3$-$1~$R_{\rm WD}$,][]{Allan1998,Hayashi2014}.
The X-ray absorption column of \gcas\ during the baseline interval is an order of magnitude smaller than that of EX Hydrae,
possibly suggesting even larger accretion spots.
The absence of an X-ray pulses from \gcas\ can also be explained
if the spot on the opposite side from us is large enough that a part of it always appears in our sight.

The accretion spots should be smaller than the diameter of the companion star,
i.e., $L_{\rm proj} \lesssim$ 0.02~\UNITSOLARRADIUS\ for a 1~\UNITSOLARMASS\ WD \citep{Provencal1998}.
The projection velocity of a blob relative to the X-ray emitting source should be
$v_{\rm proj} \sim L_{\rm proj}/T_{\rm tr} \lesssim$20~\UNITVEL\ for softness dips 1, 3, 4 and 6.
We here consider the following three components in the presumed Be+WD binary system that could work as X-ray absorbers:
i) Be stellar wind,
ii) Be decretion disk, and
iii) WD accretion disk or the accretion flow onto the companion.
For case i), 
the wind velocity is measured at $\sim$1700~\UNITVEL \citep{HenrichsP83,SmithRobinson99}.
Since the companion was almost at quadrature during the \SUZAKU\ observation ($\phi\sim$0.57 in the ephemeris in Orbit 1 in \citealt{Smith2012a}),
its projection velocity is $\sim$1500~\UNITVEL, which may be too fast for the X-ray absorbers.
For case ii),
the decretion disk rotates much slower around the primary star, but
the separation between the Be star and the 1\UNITSOLARMASS\ companion is $\sim$350\UNITSOLARRADIUS,
while the Be disk should not extend beyond the Roche lobe radius, which is $\sim$213\UNITSOLARRADIUS\ \citep{Eggleton1983}.
Also the Be disk may not be in the line of sight to the companion star.
For case iii),
the Keplerian motion is $\sim$40~\UNITVEL\ at 100~\UNITSOLARRADIUS\ from a 1~\UNITSOLARMASS\ companion, and faster at smaller distances.
The absorbers should rotate very far from the WD in an orbit oblique to the orbital plane to meet the condition $v_{\rm proj} \lesssim$20~\UNITVEL.
All these components are expected to move faster than the estimated velocity range of the blob.
We may need to consider another possible absorbing source.

We also compare the X-ray column density measurements of \gcas, both during the softness dips and the baseline interval, to that expected if the X-ray source is indeed at the position of the WD and embedded in the Be wind.  For a separation of $a=$370\UNITSOLARRADIUS, a radius of $R=$10\UNITSOLARRADIUS, and a $\beta=1$ velocity law, i.e.\ $v(r) = v_\infty(1-R/r)^\beta$, the column density of a smooth, non-rotating wind to the WD at quadrature (which is independent of the orbital inclination for a circular orbit) is \NH~$=(\dot{M}_{-7}/v_{\infty,8})\,3.1\times10^{20}$~\UNITNH\
where $\dot{M}_{-7}$ is the mass-loss rate in units of 10$^{-7}$ \UNITSOLARMASSYEAR, and $v_{\infty,8}$ is the terminal velocity of the wind $v_\infty$ in units of 10$^8$~{\rm cm~s$^{-1}$}.
Rotation lowers the wind density as a function of the polar angle to the rotation axis $\theta$ by $\rho(\theta)\sim\sqrt{1-\Omega \sin^2(\theta)}$ \citep{OwockiGayleyCranmer98} where $\Omega=v_{\rm rot}^2R/GM$ is the square of the fraction of the equatorial rotation speed $v_{\rm rot}$ to the equatorial Keplerian velocity.  As Be stars are rapid rotators, values of $\Omega=0.8$\,[0.95] have densities at the equator that are 0.45 [0.22] lower than the poles, so the quoted \NH\ formula above is an upper limit.
Alternatively, any gravitational focusing of the wind by the WD (most likely a small effect) and material flowing around the WD (a larger effect) will increase the \NH\ values obtained from this calculation.
Additionally, the inclusion of clumping can cause \NH\ to fluctuate in time by an order of magnitude depending on the clump model \citep{OskinovaFeldmeierKretschmar12}, as indeed is shown by this analysis. 
So ideally the computed \NH\ values should be between the baseline interval \NH\ $=3.8\times10^{20}$~\UNITNH\ and the softness dip interval \NH$=2.4-8.1\times10^{21}$~\UNITNH.
Plugging in the \gcas\ parameters of $v_{\infty,8}\approx1.7$ \citep{HenrichsP83,SmithRobinson99} and mass-loss rates ranging from $\dot{M}_{-7}\approx0.1$ \citep{HenrichsHammerschlag-HensbergeLamers80} to $\dot{M}_{-7}\approx5$ \citep[upper limit in][lower value quoted is $\dot{M}_{-7}\approx0.25$]{LamersWaters87}, the expected absorbing column ranges from \NH $=0.18-9.2\times10^{20}$~\UNITNH.
This upper limit lies within the softness dip interval and baseline interval values, consistent with the high mass-loss rate found by \citet{LamersWaters87}.

In either scenario, some of these absorbers may be trapped by the companion and accreted onto its surface.
The absorber mass is expressed as $M_{\rm abs} \sim \gamma m_{p} \rho_{\rm abs} l_{\rm d} l_{\rm proj}^{2} \sim \gamma m_{p}$ \NH $(L_{\rm proj} T_{\rm fb}/T_{\rm tr})^{2}$
where $m_{p}$ is the proton mass and $\gamma$ is mass fraction with respect to hydrogen (1.39 for solar abundance matter).
For softness dip 4, $M_{\rm abs} \sim$5$\times$10$^{16}$~g $\sim$3$\times$10$^{-17}$~\UNITSOLARMASS.
If a blob similar to this size accretes onto the presumed WD companion,
the released gravitational energy is $\sim$10$^{34}$~ergs,
which is comparable to the X-ray energies of the shot events \citep{Robinson2000}.

The work suggests stable X-ray emission from two plasma components.
The cool component has log~\LX/\Lbol~$\sim-$7.3 (\Lbol: bolometric luminosity), \KT~$\sim$0.9~keV and no apparent time variation,
which are similar to X-ray emission from OB stars \citep[e.g.,][]{Berghoefer1997}.
This component perhaps originates from radiatively driven winds from the primary Be star.
The hot component has only a factor of 3 higher luminosity, $\sim$2.0$\times$10$^{31}$~\UNITLUMI, than the cool component,
but it has a very high plasma temperature at \KT~$\sim$5~keV.
It may originate from the head-on collision between the Be disk and the WD wind or the primary wind
if the disk is not wind fed.

A WD is the end product of a star with an initial mass of $\lesssim$10~\UNITSOLARMASS\ \citep[see references in][]{Winget2008}.
A star born with $\sim$15~\UNITSOLARMASS\ should explode as a core-collapse supernova before the companion evolves into a WD.
A possible evolutionary channel that produces a Be + WD binary system is binary interaction.
At a late evolutionary stage, significant mass of the (current) companion may be stripped off to the (current) primary star
through mass transfer.
The companion loses mass to drive core nuclear burning and becomes a WD, while the primary star increases its mass 
to become a Be star.
Simulations suggest that binary interaction can produce 10$^{5}$ Be + WD binary systems in our galaxy \citep{Shao2014}.
The gainer is expected to become an outlier among the cluster members in the HR diagram, recognized as a blue straggler \citep{Knigge2009}.
Actually, a few \gcas\ analogues are classified as blue stragglers \citep{SmithM2016}.
The rapid rotation and Be disk of \gcas\ and its analogues could be an outcome of the mass transfer activity\footnote{
We, here, do not mention the origin of the rapid rotation and Be disk of the classical Be stars, in general.
\citet{Meurs1992} found insignificant difference in X-ray luminosity between the classical Be stars and normal OB stars,
and therefore did not suggest the presence of a compact companion for every classical Be star.}.

\citet{Mannucci2006} suggest two populations in Type~Ia supernovae (SNe): 
one explodes within 10$^{8}$~yr of their stellar birth and the other with a much wider time range up to $\sim$3~Gyr.
The \gcas\ companion could explode as a ``prompt" Type~Ia SN if the supposed mass accretion continues.
Assuming a WD companion with $M\sim$1~\UNITSOLARMASS, $r\sim$0.01~\UNITSOLARRADIUS,
and a 10\% conversion efficiency from gravitational energy to X-ray energy,
the mass accretion rate is estimated at $\approx$8$\times$10$^{-10}$~\UNITSOLARMASSYEAR.
At this rate, it takes $\approx$5$\times$10$^{8}$ years to gain $\sim$0.4~\UNITSOLARMASS\ and reach the Chandrasekhar mass limit,
which is longer than the lifetime of a 15~\UNITSOLARMASS\ star.
The mass accretion rate needs to be actually higher than the above estimate or increase significantly in a later stage
if the WD companion is to explode as a ``prompt" Type~Ia SN.

\section{Conclusion}

The \SUZAKU\ XIS observation of \gcas\ in 2011 July detects flux variations 
on short ($\lesssim$1~ksec) and long ($\gtrsim$10~ksec) timescales in all of the 0.5$-$1, 1$-$2, 2$-$4 and 4$-$9 keV bands.
These light curves look very similar at a first glance, but they show abrupt color variations, which are recognized as ``softness dips"
in time series of the softness ratios.
These observed variations are typical of \gcas\ as earlier X-ray observations detected similar flux and color variations from the star.
We identify six softness dip events in time series of the softness ratio of the 0.5$-$1~keV band over the 4$-$9 keV band.
These dips are recognized more weakly in the softness ratios of the harder energy band
and apparently uncorrelated to the 4$-$9~keV flux variation.
This result indicates that the softness dips are produced by sudden increases of X-ray absorption in the line of sight, and not the variation of the plasma temperature.

All six softness dips are symmetrical in time series of the softness ratios 
and some of them show saturation-like flat bottoms.
Each softness dip can be reproduced by a symmetrical trapezoidal shape with two transition timescales of 0.4$-$1~ksec and 4$-$7~ksec.
All spectra during each softness dip interval and outside of the dips (baseline interval) can be reproduced by \KT~$\sim$12~keV plasma emission 
with iron $K$ fluorescence.
The baseline spectrum suffers almost negligible absorption at \NH~$\sim$3.8$\times$10$^{20}$~\UNITNH, which is consistent with the optical extinction.
The softness dip spectra need additional absorption of \NH~$\sim$2.4$-$8.1 $\times$10$^{21}$~\UNITNH\ covered 
partially by $\sim$35$-$80\%.
The absorbers are perhaps blobs passing over an X-ray emitting region,
which is perhaps one of the magnetically funneled accretion spots of a WD companion orbiting around the Be star.

The softness ratio in each energy band has a weak but significant correlation with the 4$-$9 keV flux,
suggesting the presence of stable, soft X-ray emission.
The stable spectrum reproduced from the correlation plots requires two plasma components at 
\KT~$\sim$0.9 and 5.4~keV with negligible absorption, which probably correspond to the intermediate temperature
components found from high resolution grating spectra.
The cool component probably originates from the Be primary stellar wind,
while the hot component may originate from the head-on collision of either the Be or WD wind with the Be disk.

The formation of a Be + WD binary requires mass transfer between two relatively massive stars.
Gamma Cas may have experienced such activity in the past, which might be responsible for the rapid rotation of the Be star 
and the formation of the Be disk.
If significant mass accretes back from the Be star during its lifetime, the WD companion might explode as a prompt Type~Ia supernova.

\acknowledgments

The authors are grateful to the anonymous referee for very useful comments and suggestions that helped 
to significantly improve the paper.
The authors appreciate Drs.\ Vanbeveren, D., Drake S., Smith, M., Mukai, K., Pottschmidt, K., and Shrader, C.
for providing important information and discussion.
This research has made use of data obtained from the High Energy Astrophysics Science Archive
Research Center (HEASARC), provided by NASA's Goddard Space Flight Center.
This research has made use of NASA's Astrophysics Data System Bibliographic Services.
K.H. is supported by the \CHANDRA\ grant GO4-15019A, the \XMM\ grant NNX15AK62G, and 
the ADAP grant NNX15AM96G.
C.M.P.R. is supported by an appointment to the NASA Postdoctoral Program at the Goddard Space Flight Center, 
administered by Universities Space Research Association under contract with NASA.

\facility{Suzaku (XIS)}


\bibliographystyle{aasjournal}
\bibliography{inst,sci_AI,sci_JZ,scibook,ChrisBib_gCas_KH16}

\clearpage

\begin{deluxetable}{LcCCCCCC}
\tablecolumns{8}
\tablewidth{0pc}
\tabletypesize{\scriptsize}
\tablecaption{Softness Dip Parameters\label{tbl:SR_curve_fit}}
\tablehead{
\colhead{$i$}&
&
\colhead{1}&
\colhead{2}&
\colhead{3$^{{\dagger}0}$}&
\colhead{4}&
\colhead{5}&
\colhead{6}
}
\startdata
\multicolumn{8}{l}{Softness Dip}\\
~~t_{\rm c}^{i}&(ksec)&\leq13.4\pm0.5^{{\dagger}1}& 56.1\pm0.4&65.1\pm0.1&72.7\pm0.2&81.7\pm0.5&\geq109.4\pm0.8^{{\dagger}1}\\
~~T_{\rm fb}^{i}&(ksec)&\geq22.5\pm1.0^{{\dagger}1}& 0.0\pm2.2&0.0\pm0.1&2.4\pm0.7&0.1\pm3.3 &\geq7.4\pm1.7^{{\dagger}1}\\
~~T_{\rm tr}^{i}&(ksec)&0.5\pm0.5& 7.5\pm1.4&1.1\pm0.1&1.0\pm0.7&4.0\pm1.4&1.2\pm3.8\\
~~SR_{\rm d}^{i}&&0.18\pm0.01&0.22\pm0.03&0.35\pm0.03&0.33\pm0.02&0.16\pm0.09&0.11\pm0.02\\ \hline
\multicolumn{8}{l}{Baseline}\\
~~SR_{\rm bl}&&\multicolumn{6}{C}{0.61\pm0.01}\\
\enddata
\tablecomments{
The errors denote 1$\sigma$ confidence ranges.
${{\dagger}0}$: The other parameters are fixed in their error estimates.
${{\dagger}1}$: These values and errors are derived assuming that softness dip 1 (6) falls (recovers) at the start (end) of the observation.
Since the softness dip starts (ends) before (after) the observation, only these limits can be placed.
}
\end{deluxetable}

\begin{deluxetable}{lcccccccc}
\tablecolumns{9}
\tablewidth{0pc}
\tabletypesize{\scriptsize}
\tablecaption{Simultaneous Fit of the Softness Dip Spectra\label{tbl:dip_spec_pcf}}
\tablehead{
&&
\colhead{Baseline}&
\colhead{1}&
\colhead{2}&
\colhead{3}&
\colhead{4}&
\colhead{5}&
\colhead{6}
}
\startdata
\multicolumn{8}{l}{Variable Emission}\\
~~\KT[$V$]&(keV)&\multicolumn{7}{c}{12.1$^{+0.3}_{-0.1}$}\\
~~$Z$[$V$]&(solar)&\multicolumn{7}{c}{0.32$\pm$0.02}\\
~~\EM[$V_i$]&(10$^{55}$~\UNITEI)&3.1$\pm0.03$&4.9$\pm$0.04&4.1$\pm$0.05&3.8$\pm0.13$&3.5$\pm$0.05&3.2$\pm$0.06&3.5$\pm$0.05\\
~~F$_{{\rm X, FeI} K\alpha}$[$V_i$]&(10$^{-5}$ ph cm$^{-2}$ s$^{-1}$)&5.3$\pm$0.9&7.7$\pm$1.5&10.7$\pm$2.6&0.0$^{+12}$& 6.5$\pm$2.4 & 3.6$\pm$3.1&4.5$\pm$2.1\\
~~EW$_{{\rm FeI} K\alpha}$[$V_i$]&(eV)&57$\pm$9&53$\pm$10&87$\pm$21&0$^{+105}$&60$\pm$23 &37$\pm$33&42$\pm$20\\
~~\NH$_{\rm, PC}$[$V_i$]&(10$^{21}$~\UNITNH)&\nodata&5.3$^{+0.8}_{-0.7}$&4.9$^{+1.2}_{-1.1}$&5.3$^{+1.6}_{-1.4}$ &7.6$^{+0.9}_{-0.8}$&1.8$^{+1.9}_{-1.2}$&3.1$^{+1.6}_{-1.3}$\\
~~{\it CF}$_{\rm PC}$[$V_i$]&(\%)&\nodata&41$\pm$3&44$^{+6}_{-5}$&80$^{+13}_{-10}$&65$\pm$3&45$^{+54}_{-17}$ &35$^{+14}_{-7}$\\
~~\NH[$V$]&(10$^{21}$~\UNITNH)&\multicolumn{7}{c}{0.38$^{+0.04}_{-0.06}$}\\
~~\FX&(10$^{-10}$~\UNITFLUX)&1.2&1.8&1.5&1.3&1.2&1.2&1.4\\
~~\LX&(10$^{32}$~\UNITLUMI)&5.3&8.3&7.0&6.5&6.0&5.4&6.0\\
\multicolumn{8}{l}{Stable Emission}\\
~~\KT[$S_{\rm C}$]/\KT[$S_{\rm H}$]&(keV)&\multicolumn{7}{c}{0.96$^{+0.03}_{-0.04}$/5.4$_{-1.2}$}\\
~~\EM[$S_{\rm C}$]/\EM[$S_{\rm H}$]&(10$^{54}$~\UNITEI)&\multicolumn{7}{c}{0.29$^{+0.05}_{-0.03}$/1.1$^{+0.05}_{-0.28}$}\\
~~\NH[$S$]&(10$^{21}$~\UNITNH)&\multicolumn{7}{c}{0.06$^{+0.77}$}\\
~~\FX[$S_{\rm C}$]/\FX[$S_{\rm H}$]&(10$^{-12}$~\UNITFLUX)&\multicolumn{7}{c}{1.5/4.8}\\
~~\LX[$S_{\rm C}$]/\LX[$S_{\rm H}$]&(10$^{31}$~\UNITLUMI)&\multicolumn{7}{c}{0.63/2.0}\\ \hline
\multicolumn{2}{l}{Reduced $\chi^{2}$ (dof)}&\multicolumn{7}{c}{1.35 (2208)}\\
\enddata
\tablecomments{
The errors denote 90\% confidence ranges.
EW: Equivalent Width.
{\it CF}$_{\rm PC}$: Covering Fraction.
$V_{(i)}$: ($i$-th) variable component.
$S_{\rm (C/H)}$: (cool/hot) stable component.
The abundances of the stable emission components are fixed at 1 solar value.
\FX: observed flux between 0.3$-$10~keV.
\LX: absorption corrected intrinsic luminosity between 0.3$-$10~keV.
The distance is assumed at 188~pc.
}
\end{deluxetable}

\end{document}